# Learning Nonlinear Waves in Plasmon-induced Transparency

**Jiaxi Cheng,[1][*] Zhenhao Cen,[2] and Siliu Xu[3][**]**

[1]*Department of Computer Science, University of Waikato, Hamilton, New Zealand*
[2]*Shenzhen Graduate School, Peking University, Shenzhen, China*
[3]*School of Electronic and Information Engineering, Hubei University of Science and Technology, Xianning, China*
[*]*Corresponding author: jc411@students.waikato.ac.nz 1601213918@sz.pku.edu.cn*
[**]*Corresponding author: xusiliu@hbust.edu.cn xusiliu1968@163.com*

**ABSTRACT**
Plasmon-induced transparency (PIT) displays complex nonlinear dynamics that find critical phenomena in areas such as nonlinear waves. However, such a nonlinear solution depends sensitively on the selection of parameters and different potentials in the Schrödinger equation. Despite this complexity, the machine learning community has developed remarkable efficiencies in predicting complicated datasets by regression. Here, we consider a recurrent neural network (RNN) approach to predict the complex propagation of nonlinear solitons in plasmon-induced transparency metamaterial systems with applied potentials bypassing the need for analytical and numerical approaches of a guiding model. We demonstrate the success of this scheme on the prediction of the propagation of the nonlinear solitons solely from a given initial condition and potential. We prove the prominent agreement of results in simulation and prediction by long short-term memory (LSTM) artificial neural networks. The framework presented in this work opens up a new perspective for the application of RNN in quantum systems and nonlinear waves using Schrödinger-type equations, for example, the nonlinear dynamics in cold-atom systems and nonlinear fiber optics.

## 1 Introduction

Groundbreaking advances in artificial intelligence (AI) in the areas of robot technology [1], optimization for batteries [2], and achieving remarkable performance in assignments that traditional algorithms require massive time [3] have captivated public interests in recent years. The applications of the machine learning approach in the prediction of physics [4] were booming lately by the trademark prowess of neural networks (NN) in regression and classification tasks. In Schrödinger-type systems, however, the selection of guiding parameters is susceptible to the evolution of solitons. Application of traditional methods such as numerical and analytical solutions are challenging because of limited power, memory and the speed of the hardware, and the mathematical complexity of the equations. However, given that the exciting chance of utilizing machine learning as a tool to process a large amount of data in nonlinear waves and predict the soliton shapes is still largely unexplored, the spectacular promise paves the way for advances in machine learning in physics.

In recent years, manipulation of quantum mechanical properties for some quantum technologies has improved the performance of these applications. Electromagnetically induced transparency (EIT) is one example of controlling atoms by strong coupling near-resonant optical fields [5-7]. EIT has been realized and analogized by metamaterials, which are structured on subwavelength scales in recent years [8-10]. Plasmon-induced transparency (PIT) in metamaterials is a physical effect that can be analogized to EIT and is known to exhibit surface plasmonic polaritons that are tightly tied to subwavelength-periodicity pits [11, 12], which have many similarities with EIT optical solitons in cold atomic systems [13, 14]. PIT is a destructive interference effect that arises from the strong coupling between a few modes, which can be surface plasmon polaritons of the meta-atoms in metamaterials. PIT has been applied in a variety of fields, including plasmonic waveguide modulator [15], compact logic device [16], and sensing applications [17]. The nonlinear phenomena in the absorption spectrum generated in PIT metamaterial are of vast interest in nonlinear regimes since various types of solitary waves can be provided by solving the nonlinear Schrödinger equation (NLSE) analytically and numerically.

Neural networks, as one of the branches in machine learning, has applied to analyze rogue waves in supercontinuum generation [18], identify the number of solitons produced by random processes in optical wavelength-division multiplexed telecommunication systems [19], and revealing internal dynamics evolution of soliton molecules from the real-time spectral interferences [20]. Additionally, the application of mode-locked laser [21, 22] and modulation instability in optical fiber technology [23] have been clarified numerically. While these applications have used machine learning techniques, the prediction methods are limited to feedforward neural networks, the architecture of which is relatively simple. In our work, regression as one of three areas of machine learning is utilized to predict the evolution of the solitons by creating RNN models, which is one of the knowledge-based methods that perform well in nonlinear dynamics, for example, high-dimensional chaotic systems [24].

More generally, experiments and theoretical work in nonlinear dynamics with EIT and PIT are of vast attraction since the potential applications [25] in optical soliton formation as a primary phenomenon common to lots of Schrödinger-type systems [26, 27]. However, due to the complexities of the problems of limitations of the computer hardware, the design and analysis of experiments in EIT and PIT require massive time in analytical and numerical calculations to generate specific types of solitons in line with simulations and approximate solutions. The computationally demanding dilemma facilitates the unprecedented booming of artificial intelligence since the traditional methods face a bottleneck in modern times.

Recently, the simulation of the nonlinear Schrödinger equation (NLSE) in PIT has attracted extensive interest since new types of solitons have been predicted and observed [28]. However, in most cases, analytical methods cannot acquire the exact solutions of a Schrödinger-type system. Due to the relationships between the numerical results and the initial solutions, and the parameters, the system design requires massive computing time to integrate the equations. This phenomenon creates a barrier to the application of numerical methods in our scheme.

In this paper, the authors first propose the (2+1) dimensional solitons in PIT to analyze the results in conditions in which parameters are sensitive to propagate solitons in NLSE. The RNN architecture was constructed to solve the numerical problems in PIT metamaterial systems with specific potentials, bypassing the need for complex mathematical derivation and demanding algorithms to acquire the solutions of the governing propagation model. In our cases, two forms of potentials concerning soliton types have been studied by exploiting the learning efficiency of long short-term memory (LSTM) recurrent neural networks in the evolution of solitons of PIT metamaterial with external potentials through comparison with numerical solutions. We prove that there is a definitive agreement with the predicted and simulated results.

## 2 Model

In this article, the metamaterial is used to generate PIT, which was experimentally demonstrated in Zhu et al.'s work [29]. Here, a periodic array of unit cells that can be treated as magnetic meta-atoms with a pair of split-ring resonators inserted with double varactors (SRR) is used, which is presented in Figure 1(a). The length of each resonator is *l = 8 mm*, the thickness of the aluminum strips *w = 0.5 mm*, and the gap size *h = 0.25 mm* to generate a magnetic response for the resonator [30]. The resonance frequencies of two SRRs are hypothesized the same for simplicity in RLC circuit analog, including bright oscillators and dark oscillators corresponding to the left and right loops. The capacitance, resistance, and inductance, which are C, R, and L in Methods, along with electromotive voltage, provide a convenient means of studying and calculating the derivation of the coupled nonlinear Schrödinger equation. The coupled equations of renormalized voltage $q_1$ and $q_2$ can be described using Kirchhoff's law (see Methods). As illustrated in Figure 1(c), the incident electromagnetic fields of the framework of metamaterial are described in Z and XY directions since the magnetic effects can be produced in this selection. Figure 1(b) shows the relationships between absorption spectrum and coupling strength with three cases in which PIT effect achieve in PIT regime for the separations between SRRs *d=4.5 mm*.

The evolution of solitons in NLSE can be represented as a sequence of real and imaginary parts of amplitude distributions at points along the propagation axis. The value of intensity that equals the nonnegative square root of real and imaginary parts is determined by the propagation that precedes it and the generation point, which conventionally requires a great deal of time by a great many simulation steps for NLSE.

For simplicity and time, we propose a model-free method called RNN instead of the traditional approach. RNN is a specific type of neural network which possesses internal memory, allowing them to account for long-term dependencies and thus to identify patterns in sequential data [31] robustly. A significant application of RNN is to predict time-series data since it has adapted explicitly to model dynamic behaviors [3] lately, including sequence generation [32], feature extraction in cybersecurity [33], and active distribution systems [34]. Equally, it can be applied to predict the evolution of optical solitons in PIT metamaterial systems naturally with various potentials.

The schematic of RNN is presented in Figure 1(e), and the model of RNN is shown in Figure 1(d) with four structures: input layer, LSTM layer, dense layer, and the output layer. We provide input data from the input layer for the machine to learn, and in our work, it is the values of intensity of solitons in a three-dimensional matrix. Specifically, the LSTM layer uses the current and the memory states to predict the intensity profile at the specified point by a feedback loop.

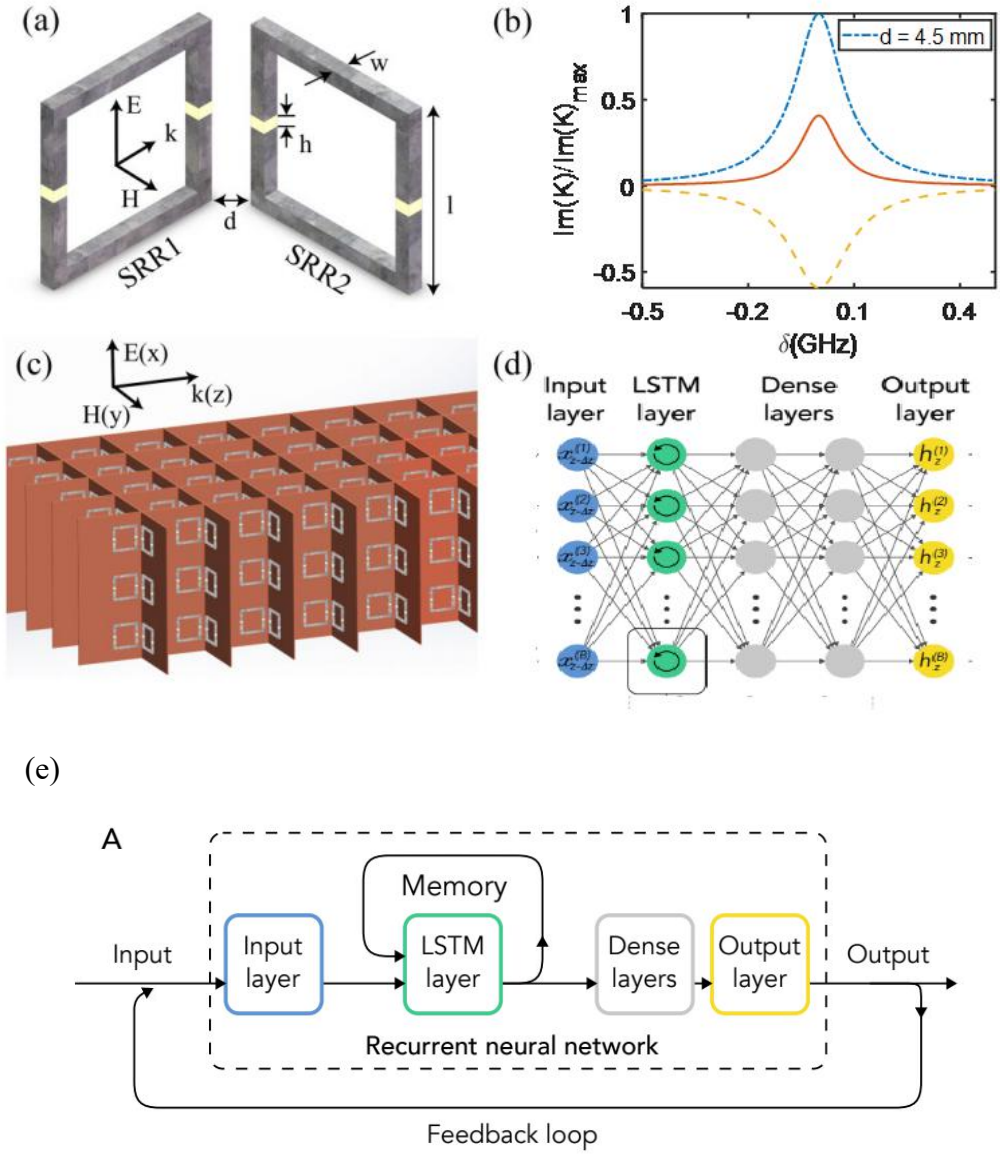

FIG. 1 (a) The metamaterial element containing two SRRs with *l*=8.0 *mm*, *w*=0.50 *mm,* and *h*=0.20 *mm*. The directions of the radiation field, magnetic field, and electric field are set to be mutually perpendicular. SRR1 generates bright soliton, whereas the other generate dark soliton. (b) The PIT effect in metamaterial array in PIT regime with *d=4.5 mm* by providing absorption spectrum and coupling strength. (c) Schematic of the periodic array of metamaterial elements with the inlet in k direction, electrical field intensity, and magnetic field intensity. (de) The model of RNN architecture shows the four layers: the input layer, the LSTM layer, the dense layer, and the output layer.

3.0 **Numerical Simulation**
Before going into the details of the evolution of the solitons in the PIT metamaterial system, we firstly introduce that it is governed by Maxwell's equation. After the second and third-order approximation [35] acquired by the multiple scales method [36], we neglect the impact of diffraction effects on x directions based on the assumption of the nonlinear waves in (2+1) dimensions.

In Figure 2, we carry out the simulation method of employment of the applied potential with $V = 2.8\delta[\sin(2x)+\sin(2t)]$ NLSE, where $\delta$ forms the normalization parameter. A magnetic soliton is generated in the PIT metamaterial for the normalized intensity with the function of x and time (Figure 2(a)) with the applied potential. The detailed propagation of ultraslow magnetic soliton concerning time has been provided in Figure 2(b). These two figures plot the evolution of intensity profile in the metamaterial, which the type of the soliton is single peak one. In Figure 2(c), we present the result of numerical simulations that illustrate the normalized peak power (P) for the various amplitude $V_0$ of potential $V = \delta V_0[\sin(2x)+\sin(2t)]$. Our numerical results deduce that the peak power P has a nearly linear dependence relation with $V_0$ being varied from 2.5 to 3.2. One can see the dependence of the amplitude of applied potential $V_0$ on the chemical potential $\mu$. The curve in Figure 2(d) is similar to a linear one.

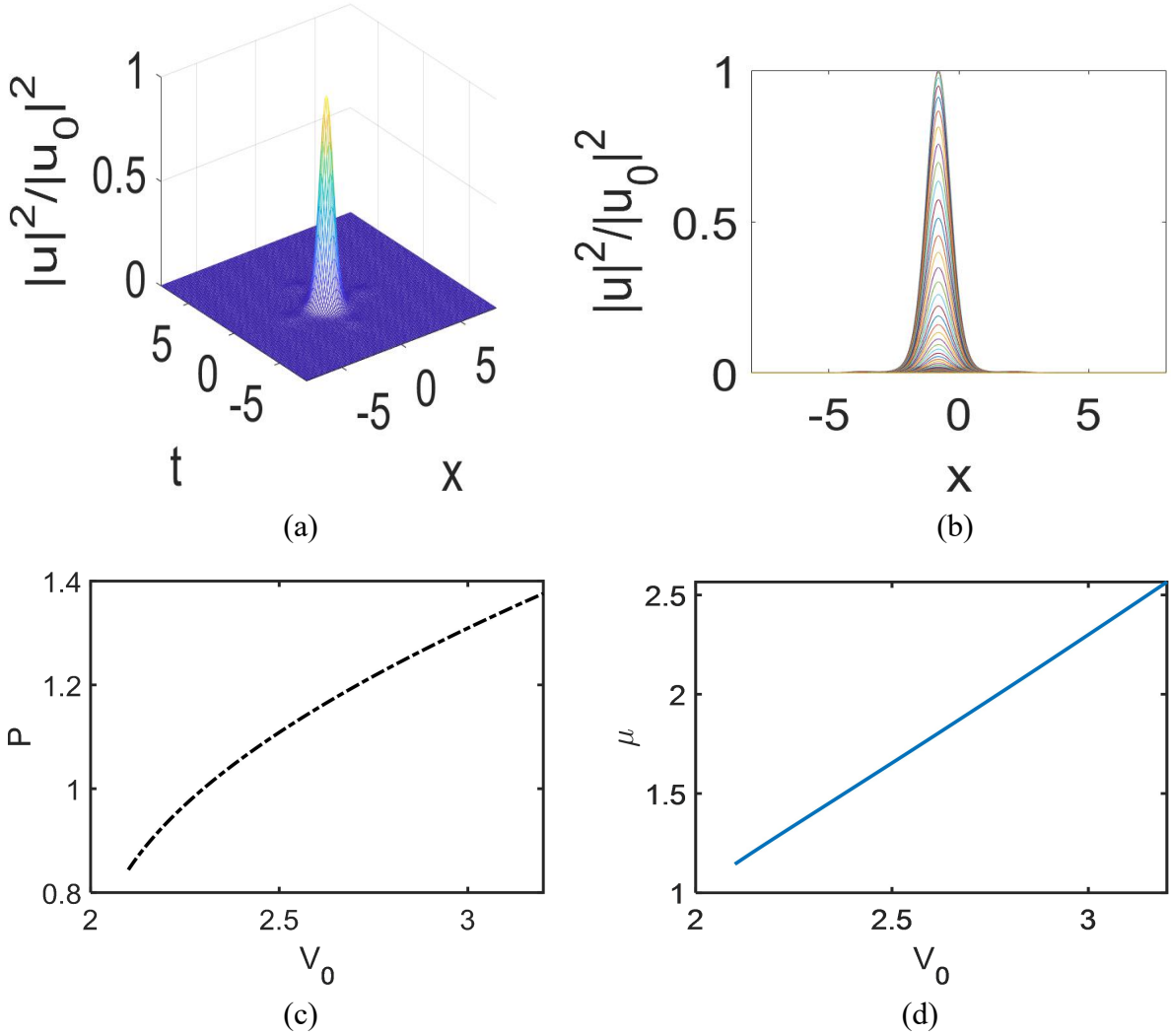

(a) (b)
(c) (d)

FIG. 2 (a-b) Intensity of the wave function distribution of the propagation of the ultraslow magnetic solitons in the PIT metamaterial with potential $V = \delta V_0[\sin(2x)+\sin(2t)]$. (a) illustrates the profile of the solitons in the three-dimensional diagram, and (b) shows the intensity in the x-direction at selected distances. (c-d) Simulated absolute normalized peak power (P) and chemical potential ($\mu$) for various $V_0$.

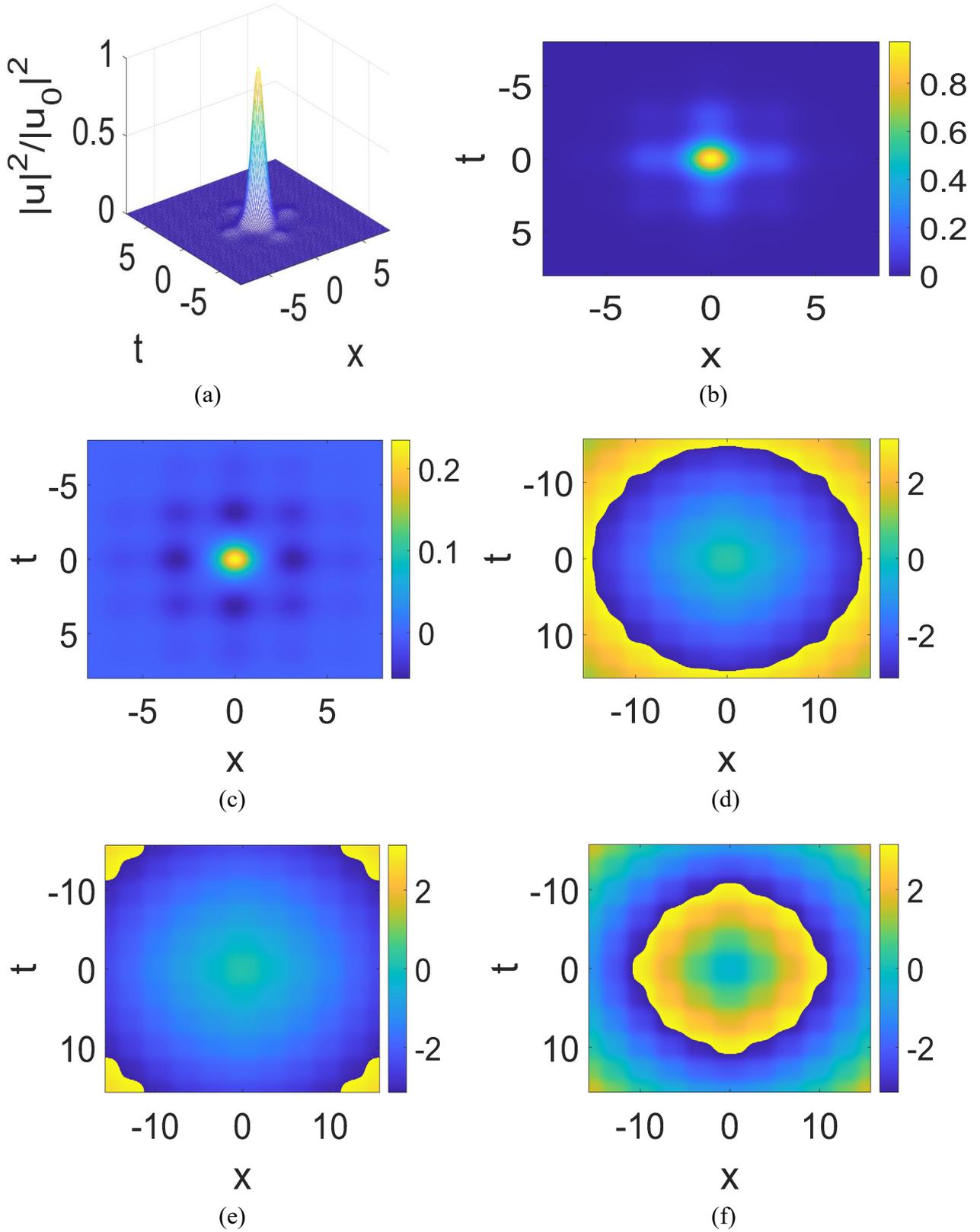

FIG. 3 The simulation results of NLSE in PIT metamaterial with potential $V = \delta V_0 [\cos^2(x) + \cos^2(t)]$ with $V_0 = 3.2$, $V_0 = 3.18$, and $V_0 = 3.26$ corresponding to (a)-(d), (e), and (f), (a) Plot showing the intensity distribution of the nonlinear waves as a periodic potential. (b)-(c) The wave function distribution for the real and imaginary part in the X-Y plane. (d)-(f) Phase angle about the wave function in PIT metamaterial.

We now examine the simulation results of another potential $V = \delta V_0[\cos^2(x) + \cos^2(t)]$ in the PIT metamaterial array, with the definite values of $V_0$. Results are shown in Figure 3 for $V_0 = 3.2$, $V_0 = 3.18$, and $V_0 = 3.26$ associated with (a)-(d), (e), and (f), respectively. These values are chosen as they lead to a single-pole output soliton with very distinct characteristics. Figure 4(a) shows the intensity of the normalized wave distribution and the evolution of the magnetic solitons in the PIT metamaterial array. For convenient visualization, the propagation is plotted on a linear scale. We can see a very similar soliton shape by the intensity with Figure 2. The real and the imaginary part of wave function distribution in the X-Y plane is performed in Figure 3(b) and Figure 3(c), respectively. Figure 3(d) provides the phase angle connecting to the intensity of wave function, in this case, concluding that the value of phase angle is relatively lower centering around the peak of the wave. We also plot the phase angle $V_0 = 3.18$ following and $V_0 = 3.26$ in Figure 3(e) and Figure 3(f), respectively. In contrast, the values of phase angle larger than 2 occur in the four corners for the first example, whereas it is around the solitons for the second one.

**5 RNN Results**

We tested the RNN for complete evolution predictions in the time domain of solitons with PIT metamaterial. One of the best performances of NN is observed and shown in the following. The machine learning results for the cases corresponding to Figure 2 and Figure 3 are shown in Figure 4. We illustrate the results obtained when training the network to model the temporal intensity propagation. It compares the propagation of the temporal intensity simulated using NLSE in PIT metamaterial (left panel) with the RNN prediction results (central panel). The RNN results in Figure 4(a) and (b) correspond to the potential $V = \delta V_0[\sin(2x) + \sin(2t)]$ and $V = \delta V_0[\cos^2(x) + \cos^2(t)]$, respectively. It can be concluded that an excellent agreement between the numerical results and the prediction from RNN. The right panel provides the difference between discussed maps in two circumstances. One can see that the maximum difference happens at the first row, which corresponds to the potential in the form of $V = \delta V_0[\sin(2x) + \sin(2t)]$.

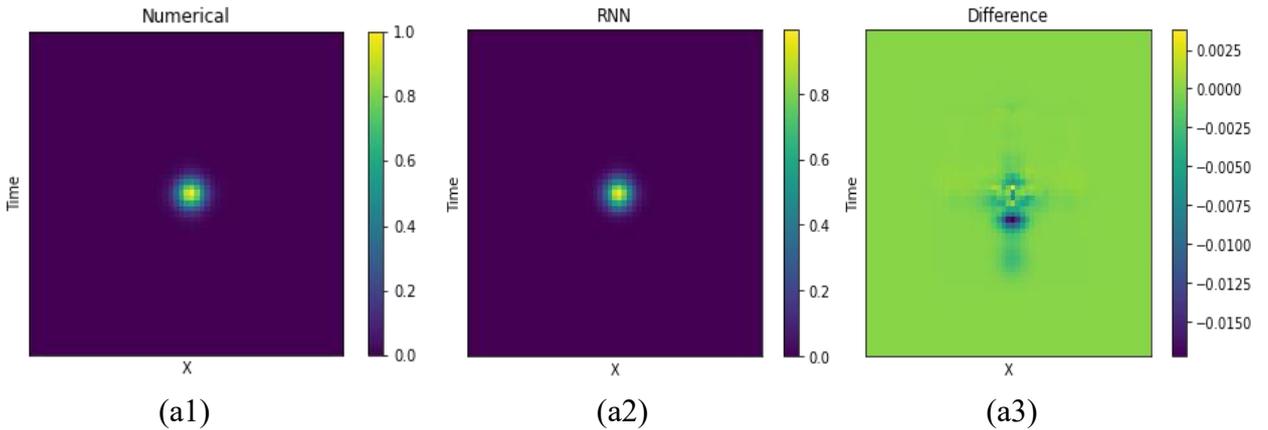

(a1)          (a2)          (a3)

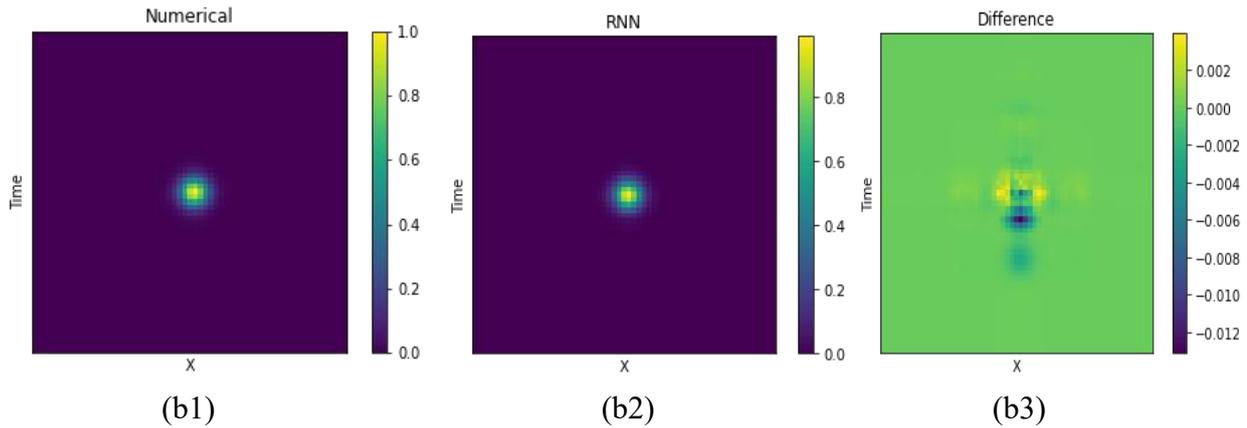

(b1)             (b2)             (b3)

FIG. 4 Temporal intensity evolution of nonlinear waves with potential $V = \delta V_0[\sin(2x) + \sin(2t)]$ and $V = \delta V_0[\cos^2(x) + \cos^2(t)]$ about (a1)-(a3) and (b1)-(b3), respectively. The panels show the numerical results (left), RNN prediction (middle), and relative difference (right).

## 8 Conclusion

We have predicted (2+1) dimensional nonlinear waves in the PIT metamaterial array with nonlinear potential applied in the system. Nonlinear dynamics of the solitons have been analyzed in our study. Specifically, we have demonstrated that the machine learning method by RNN with LSTM can excellently predict the dynamics of the single-pole solitons generated by NLSE with two forms of potentials. We confirm that the predicted results by RNN are in excellent agreement with the simulation. The machine learning results in our research prove that there are significant applications for RNN in ultrafast dynamics.

In terms of applications, the authors expect a booming period for machine learning as a tool in physics and engineering, for predicting the characteristics of the solutions of the equations, as well as the design and analysis of the experiments. In future work, the RNN can be extended in predicting the nonlinear dynamics of the waves in EIT and improve the RNN model. From a broader perspective, it is expected that RNN benefits both experiments and simulations in physics and mathematics, e.g., designing and analyzing optical experiments and predicting dynamics of nonlinear waves in Schrödinger-type systems.